\documentstyle[epsf]{article} 
\begin{document}
\title{Geometric structure of the generic \\
       static traversable wormhole throat}
\author{
David Hochberg$^{+}$ and Matt Visser$^{++}$\\
\\
$^{+}$Laboratorio de Astrof\'{\i}sica Espacial y F\'{\i}sica Fundamental\\
Apartado 50727, 28080 Madrid, Spain\\
\\
$^{++}$Physics Department, Washington University\\
Saint Louis, Missouri 63130-4899, USA\\
}
\date{gr-qc/9704082; 29 April 1997}
\maketitle
\begin{abstract}
Traversable wormholes have traditionally been viewed as intrinsically
topological entities in some multiply connected spacetime. Here, we
show that topology is too limited a tool to accurately characterize a
generic traversable wormhole: in general one needs geometric
information to detect the presence of a wormhole, or more precisely to
locate the wormhole throat. For an arbitrary static spacetime we shall
define the wormhole throat in terms of a $2$--dimensional
constant-time hypersurface of minimal area.  (Zero trace for the
extrinsic curvature plus a ``flare--out'' condition.) This enables us
to severely constrain the geometry of spacetime at the wormhole throat
and to derive generalized theorems regarding violations of the energy
conditions---theorems that do not involve geodesic averaging but
nevertheless apply to situations much more general than the
spherically symmetric Morris--Thorne traversable wormhole.  [For
example: the null energy condition (NEC), when suitably weighted and
integrated over the wormhole throat, must be violated.] The major
technical limitation of the current approach is that we work in a
static spacetime---this is already a quite rich and complicated
system.
\end{abstract}

PACS: 04.40.-b;  04.20.Cv; 04.20.Gz; 04.90.+e

e-mail: hochberg@laeff.eas.es; visser@kiwi.wustl.edu

\newpage

\section{Introduction}
\def\tr{\hbox{\rm tr}}
\def\implies{\Rightarrow}
\def\conv{\hbox{\rm conv}}
\def\Re{ {\cal R} }

Traversable wormholes~\cite{Morris-Thorne,MTY,Visser} are often viewed
as intrinsically topological objects, occurring only in multiply
connected spacetimes. Indeed, the Morris--Thorne class of
inter-universe traversable wormholes is even more restricted,
requiring both exact spherical symmetry and the existence of two
asymptotically flat regions in the spacetime.  To deal with
intra-universe traversable wormholes, the Morris--Thorne analysis must
be subjected to an approximation procedure wherein the two ends of the
wormhole are distorted and forced to reside in the same asymptotically
flat region. The existence of one or more asymptotically flat regions
is an essential ingredient of the Morris--Thorne
approach~\cite{Morris-Thorne}.

However, there are many other classes of geometries that one might
still quite reasonably want to classify as wormholes, that either
do not possess any asymptotically flat region~\cite{HPS}, or have
trivial topology~\cite{Visser}, or exhibit both these phenomena.

\begin{figure}[htb]
\vbox{\hfil\epsfbox{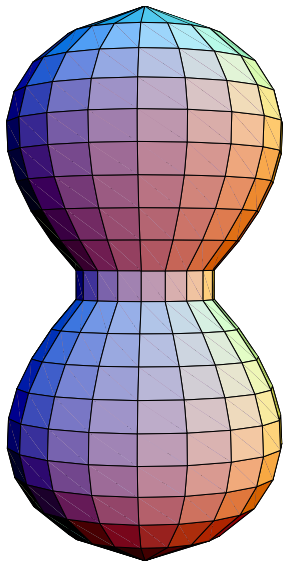}\hfil}
\caption[Dumbbell wormhole]
{\label{F-FRW}
A ``dumbbell wormhole'': Formed (for example) by two closed
Friedman--Robertson--Walker spacetimes connected by a narrow neck.}
\end{figure}

\begin{figure}[htb]
\vbox{\hfil\epsfbox{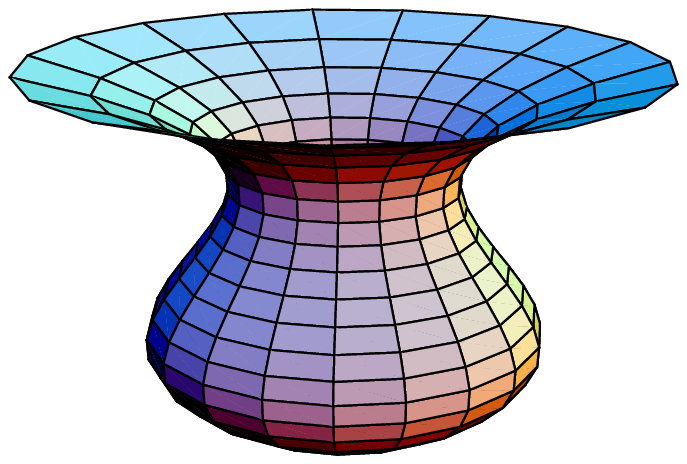}\hfil}
\caption[Topologically trivial wormhole]
{\label{F-trivial}
A wormhole with trivial topology: Formed (for example) by connecting
a single closed Friedman--Robertson--Walker spacetime to Minkowski
space by a narrow neck.}
\end{figure}

A simple example of a wormhole lacking an asymptotically flat region
is two closed Friedman--Robertson--Walker spacetimes connected by
a narrow neck (see figure~\ref{F-FRW}), you might want to call this
a ``dumbbell wormhole''. A simple example of a wormhole with trivial
topology is a single closed Friedman--Robertson--Walker spacetime
connected by a narrow neck to ordinary Minkowski space (see
figure~\ref{F-trivial}).  A general taxonomy of wormhole exemplars
may be found in~\cite[pages 89--93]{Visser}, and discussions of
wormholes with trivial topology may also be found in~\cite[pages
53--74]{Visser}.

While the restricted viewpoint based on the Morris--Thorne analysis
is acceptable for an initial discussion, the Morris--Thorne approach
fails to capture the essence of large classes of wormholes
that do not satisfy their simplifying assumptions.

In this paper we shall investigate the generic static traversable
wormhole.  We make no assumptions about spherical symmetry (or
axial symmetry, or even ``exchange'' symmetry), and we make no
assumptions about the existence of any asymptotically flat region.
We first have to define exactly what we mean by a wormhole---we
find that there is a nice {\em geometrical\,} (not topological)
characterization of the existence of, and location of, a wormhole
``throat''. This characterization is developed in terms of a
hypersurface of minimal area, subject to a ``flare--out'' condition
that generalizes that of the Morris--Thorne analysis.

With this definition in place, we can develop a number of theorems
about the existence of ``exotic matter'' at the wormhole throat.
These theorems generalize the original Morris--Thorne result by
showing that the null energy condition (NEC) is generically violated
at some points on or near the two-dimensional surface comprising
the wormhole throat.  These results should be viewed as complementary
to the topological censorship theorem~\cite{Topological-censorship}.
The topological censorship theorem tells us that in a spacetime
containing a traversable wormhole the averaged null energy condition
must be violated along at least some (not all) null geodesics, but
the theorem provides very limited information on where these
violations occur. The analysis of this paper shows that some of
these violations of the energy conditions are concentrated in the
expected place:  on or near the throat of the wormhole. The present
analysis, because it is purely local, also does not need the many
technical assumptions about asymptotic flatness, future and past
null infinities, and global hyperbolicity that are needed as
ingredients for the topological censorship
theorem~\cite{Topological-censorship}.

The key simplifying assumption in the present analysis is that of
taking a static wormhole. While we believe that a generalization
to dynamic wormholes is possible, the situation becomes technically
much more complex and and one is rapidly lost in an impenetrable
thicket of definitional subtleties and formalism.

\section{Static spacetimes}

In any static spacetime one can decompose the spacetime metric
into block diagonal form~\cite{MTW,Hawking-Ellis,Wald}:

\begin{eqnarray}
ds^2 &=& g_{\mu\nu} \; dx^\mu dx^\nu
\\
&=&
- \exp(2\phi) dt^2 + g_{ij} \; dx^i dx^j.
\end{eqnarray}

\noindent
{\em Notation:} Greek indices run from 0--3 and refer to space-time;
latin indices from the middle of the alphabet ($i$, $j$, $k$, \dots) run
from 1--3 and refer to space; latin indices from the beginning of
the alphabet ($a$, $b$, $c$, \dots) will run from 1--2 and will be used
to refer to the wormhole throat and directions parallel to the
wormhole throat.

Being static tightly constrains the space-time geometry in terms
of the three-geometry of space on a constant time slice, and the
manner in which this three-geometry is embedded into the spacetime.
For example, from \cite[page 518]{MTW} we have

\begin{eqnarray}
{}^{(3+1)}R_{ijkl} &=& 
{}^{(3)}R_{ijkl}. 
\\
{}^{(3+1)}R_{\hat tabc} &=& 0.
\\
{}^{(3+1)}R_{\hat ti\hat tj}  &=&  
\phi_{|ij} + \phi_{|i} \; \phi_{|j}.
\end{eqnarray}

\noindent
The hat on the $t$ index indicates that we are looking at components
in the normalized $t$ direction

\begin{equation}
X_{\hat t} = X_t \sqrt{-g^{tt}} = X_t \exp(-\phi).
\end{equation}

\noindent
This means we are using an orthonormal basis attached to the fiducial
observers (FIDOS). We use $X_{;\alpha}$ to denote a space-time
covariant derivative; $X_{|i}$ to denote a three-space covariant
derivative, and will shortly use $X_{:a}$ to denote two-space
covariant derivatives taken on the wormhole throat itself.

Now taking suitable contractions,

\begin{eqnarray}
{}^{(3+1)}R_{ij} &=& 
{}^{(3)}R_{ij} -  \phi_{|ij} - \phi_{|i} \; \phi_{|j}.
\\
{}^{(3+1)}R_{\hat t i} &=&  0 .
\\
{}^{(3+1)}R_{\hat t\hat t} &=&  g^{ij}
\left[  
\phi_{|ij} + \phi_{|i} \phi_{|j}
\right].
\end{eqnarray}

\noindent
So

\begin{equation}
{}^{(3+1)}R =  {}^{(3)}R - 2 g^{ij}
\left[ \phi_{|ij} + \phi_{|i} \phi_{|j} \right].
\end{equation}

\noindent
To effect these contractions, we make use of the decomposition
of the spacetime metric in terms of the spatial three-metric, the
set of vectors ${e^{\mu}_i}$ tangent to the time-slice, and the
vector $V^\mu = \exp[\phi] \; ({\partial}/{\partial t } )^{\mu}$
normal to the time slice: 

\begin{equation}
{}^{(3+1)}g^{\mu \nu} = e^{\mu}_i e^{\nu}_j \; g^{ij} - V^\mu V^\nu.
\end{equation}

Finally, for the spacetime Einstein tensor (see~\cite[page
552]{MTW})

\begin{eqnarray}
\label{E-static-stress-energy-b}
{}^{(3+1)}G_{ij} &=&  {}^{(3)}G_{ij}
 -  \phi_{|ij} - \phi_{|i} \; \phi_{|j}
+ g_{ij} \; g^{kl}\left[ \phi_{|kl} +\phi_{|k} \phi_{|l} \right].
\\
{}^{(3+1)}G_{\hat t i} &=& 0 .
\\
{}^{(3+1)} G_{\hat t\hat t} &=& + {1\over2} {}^{(3)} R.
\label{E-static-stress-energy-e}
\end{eqnarray}

\noindent
This decomposition is generic to {\em any} static spacetime.  (You
can check this decomposition against various standard textbooks to
make sure the coefficients are correct. For instance see Synge~\cite[page
339]{Synge}, Fock~\cite{Fock}, or
Adler--Bazin--Schiffer~\cite{Adler-Bazin-Schiffer})

{\em Observation:} Suppose the strong energy condition (SEC) holds
then~\cite{Visser}

\begin{eqnarray}
\label{E-static-SEC-b}
SEC &\implies& (\rho + g_{ij} T^{ij}) \geq 0
\\
&\implies& g^{ij}
\left[  
\phi_{|ij} + \phi_{|i} \phi_{|j}
\right] \geq 0
\\
&\implies& \hbox{$\phi$ has no isolated maxima}.
\label{E-static-SEC-e}
\end{eqnarray}

\section{Definition of a generic static throat}

We define a traversable wormhole throat, $\Sigma$, to be a $2$--dimensional
hypersurface of {\em minimal} area taken in one of the constant-time
spatial slices. Compute the area by taking

\begin{equation}
A(\Sigma) = \int \sqrt{{}^{(2)}g} \; d^{2} x.
\end{equation}

\noindent
Now use Gaussian normal coordinates, $x^i=(x^a;n)$, wherein the
hypersurface $\Sigma$ is taken to lie at $n=0$, so that

\begin{equation}
{}^{(3)}g_{ij} \; dx^i dx^j  = {}^{(2)}g_{ab} \; dx^a dx^b  +  dn^2.
\end{equation}

\noindent
The variation in surface area, obtained by pushing the surface $n=0$
out to $n = \delta n(x)$, is given by the standard computation

\begin{equation}
\delta A(\Sigma) = 
\int {\partial\sqrt{{}^{(2)}g}\over \partial n} \; 
\delta n(x) \; d^{2} x.
\end{equation}

\noindent
Which implies

\begin{equation}
\delta A(\Sigma) =
\int \sqrt{{}^{(2)}g} \;
{1\over2} \; g^{ab} \; {\partial g_{ab} \over \partial n} \; 
\delta n(x) \; d^{2} x.
\end{equation}

\noindent
In Gaussian normal coordinates the extrinsic curvature can be simply
defined by

\begin{equation}
K_{ab} = - {1\over2} {\partial g_{ab} \over \partial n}.
\end{equation}

\noindent
(See \cite[page 552]{MTW}. We use MTW sign conventions. The convention
in \cite[page 156]{Visser} is opposite.) Thus

\begin{equation}
\delta A(\Sigma) =  - \int \sqrt{{}^{(2)}g} \; \tr(K) \; 
\delta n(x) \; d^{2} x.
\end{equation}

\noindent
[We use the notation $\tr(X)$ to denote $g^{ab} \; X_{ab}$.] Since
this is to vanish for arbitrary $\delta n(x)$, the condition that
the area be {\em extremal} is simply $\tr(K)=0$.  To force the area
to be {\em minimal} requires (at the very least) the additional
constraint $\delta^2 A(\Sigma) \geq 0$. (We shall also consider
higher-order constraints below.)  But by explicit calculation

\begin{equation}
\delta^2 A(\Sigma) =  
- \int \sqrt{{}^{(2)}g} \;
\left( {\partial\tr(K)\over \partial n} - \tr(K)^2 \right) \; 
\delta n(x) \; \delta n(x) \; d^{2} x.
\end{equation}

\noindent
Extremality [$\tr(K)=0$] reduces this minimality constraint to

\begin{equation}
\delta^2 A(\Sigma) =  
- \int \sqrt{{}^{(2)}g} \;
\left( {\partial\tr(K)\over \partial n} \right) \; 
\delta n(x) \; \delta n(x) \; d^{2} x \geq 0.
\end{equation}

\noindent
Since this is to hold for arbitrary $\delta n(x)$ this implies that
at the throat we certainly require

\begin{equation}
{\partial\tr(K)\over \partial n} \leq 0.
\end{equation}

\noindent
This is the generalization of the Morris--Thorne ``flare-out''
condition to arbitrary static wormhole throats.

In the following definitions, the two-surface referred to is
understood to be embedded in a three-dimensional space, so that
the concept of its extrinsic curvature (relative to that embedding
space) makes sense.

\noindent
{\bf Definition: Simple flare-out condition.}\\ 
{\em A two-surface satisfies the ``simple flare-out'' condition if and
only if it is extremal, $\tr(K)=0$, and also satisfies
${\partial\tr(K)/\partial n} \leq 0$.}

This flare-out condition can be rephrased as follows: We have as
an identity that

\begin{equation}
{\partial\tr(K)\over \partial n} = 
\tr\left({\partial K\over \partial n}\right) + 2 \tr(K^2).
\end{equation}

\noindent
So minimality implies

\begin{equation}
\tr\left({\partial K\over \partial n}\right) + 2 \tr(K^2) \leq 0.
\end{equation}

\begin{figure}[htb]
\vbox{\hfil\epsfbox{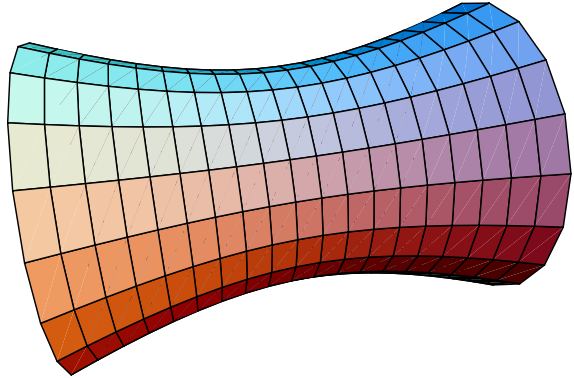}\hfil}
\caption[Generic wormhole throat]
{\label{F-generic}
Generically, we define the throat to be located at a true minimum
of the area. The geometry should flare-out on either side of the
throat, but we make no commitment to the existence of any asymptotically
flat region.}
\end{figure}

\noindent
Generically we would expect the inequality to be strict, in the
sense that ${\partial\tr(K) / \partial n} < 0$, for at least some
points on the throat. (See figure~\ref{F-generic}.) This suggests
the modified definition below.

\noindent
{\bf Definition: Strong flare-out condition.}\\
{\em A two surface satisfies the ``strong flare-out'' condition
at the point $x$ if and only if it is extremal, $\tr(K)=0$,
everywhere satisfies ${\partial\tr(K)/\partial n} \leq 0$, and 
if at the point $x$ on the surface the inequality is strict:}

\begin{equation}
{\partial\tr(K)\over\partial n} < 0.
\end{equation}

It is sometimes sufficient to demand a weak integrated form of the
flare-out condition. 

\noindent
{\bf Definition: Weak flare-out condition.}\\ 
{\em A two surface satisfies the ``weak flare-out'' condition if
and only if it is extremal, $\tr(K)=0$, and}

\begin{equation}
\int \sqrt{{}^{(2)}g} \;{\partial\tr(K)\over \partial n} d^2 x < 0.
\end{equation}

Note that the strong flare-out condition implies both the simple
flare-out condition and the weak flare-out condition, but that the
simple flare-out condition does not necessarily imply the weak
flare-out condition. (The integral could be zero.) Whenever
we do not specifically specify the type of flare-out condition
being used we deem it to be the simple flare-out condition.

\begin{figure}[htb]
\vbox{\hfil\epsfbox{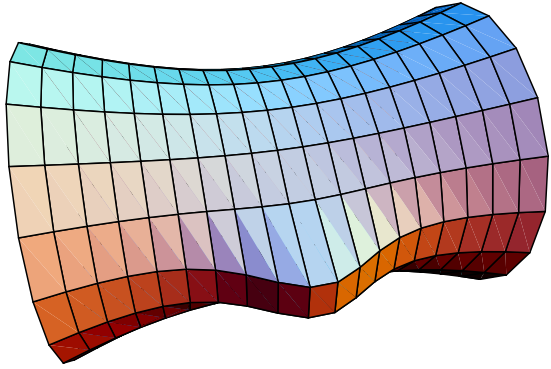}\hfil}
\caption[Strong versus weak throats]
{\label{F-bump}
Strong versus weak throats: Placing a small bump on a strong throat
typically causes it to tri-furcate into two strong throats plus a
weak throat.}
\end{figure}

The conditions under which the weak definition of flare-out are
appropriate arise, for instance, when one takes a Morris--Thorne
traversable wormhole (which is symmetric under interchange of the
two universes it connects) and distorts the geometry by placing a
small bump on the original throat. (See figure~\ref{F-bump}.)

The presence of the bump causes the old throat to trifurcate into
three extremal surfaces: Two minimal surfaces are formed, one on
each side of the old throat, (these are minimal in the strong sense
previously discussed), while the surface of symmetry between the
two universes, though by construction still extremal, is no longer
minimal in the strict sense. However, the surface of symmetry is
often (but not always) minimal in the weak (integrated) sense
indicated above. 

A second situation in which the distinction between strong and weak
throats is important is in the cut-and-paste construction for
traversable wormholes~\cite{Visser,Visser89a,Visser89b}. In this
construction one takes two (static) spacetimes (${\cal M}_1$, ${\cal
M}_2$) and excises two geometrically identical regions of the form
$\Omega_i\times\Re$, $\Omega_i$ being compact spacelike surfaces
with boundary and $\Re$ indicating the time direction. One then
identifies the two boundaries $\partial\Omega_i \times \Re$ thereby
obtaining a single manifold (${\cal M}_1 \# {\cal M}_2$) that
contains a wormhole joining the two regions ${\cal M}_i - \Omega_i
\times \Re$.  We would like to interpret the junction
${\partial\Omega_{1=2} \times \Re}$ as the throat of the wormhole.

If the sets $\Omega_i$ are convex, then there is absolutely no
problem: the junction ${\partial\Omega_{1=2} \times \Re}$ is by
construction a wormhole throat in the strong sense enunciated above.

On the other hand, if the $\Omega_i$ are concave, then it is
straightforward to convince oneself that the junction $\partial\Omega_{1=2}
\times \Re$ is not a wormhole throat in the strong sense.  If one
denotes the {\em convex hull} of $\Omega_i$ by $\conv(\Omega_i)$
then the {\em two} regions $\partial[\conv(\Omega_i)] \times \Re$
{\em are} wormhole throats in the strong sense. The junction
${\partial\Omega_{1=2} \times \Re}$ is at best a wormhole throat
in the weak sense.

For these reasons it is useful to have this notion of a weak throat
available as an alternative definition.  Whenever we do not qualify
the notion of wormhole throat it will refer to a throat in the simple
sense. Whenever we refer to a throat in the weak sense or strong
senses we will explicitly say so. Finally, it is also useful to define

\noindent
{\bf Definition: Weak $f$-weighted flare-out condition}\\ 
{\em A two surface satisfies the ``weak $f$-weighted flare-out'' condition
if and only if it is extremal, $\tr(K)=0$, and}

\begin{equation}
\int \sqrt{{}^{(2)}g} \; f(x) \; 
{\partial\tr(K)\over \partial n} d^2 x < 0.
\end{equation}

\noindent
(We will only be interested in this condition for $f(x)$ some positive
function defined over the wormhole throat.)

The constraints on the extrinsic curvature embodied in these various
definitions lead to constraints on the spacetime geometry, and
consequently constraints on the stress-energy.

\noindent
{\bf Technical point I: Degenerate throats}\\
A class of wormholes for which we have to extend these
definitions arises when the wormhole throat possesses an accidental
degeneracy in the extrinsic curvature at the throat. The previous
discussion has tacitly been assuming that near the throat we can
write

\begin{eqnarray}
{}^{(2)} g_{ab}(x,n) &=& 
{}^{(2)} g_{ab}(x,0) + 
n \left.{\partial\left[{}^{(2)} g_{ab}(x,n)\right]
\over \partial n}\right|_{n=0} 
\nonumber\\
&& \qquad
+ {n^2\over2} 
\left.{\partial^2\left[{}^{(2)} g_{ab}(x,n)\right]
\over (\partial n)^2}\right|_{n=0} +
O[n^3].
\end{eqnarray}

\noindent
with the linear term having trace zero (to satisfy extremality) and the
quadratic term being constrained by the flare-out conditions. 

Now if we have an accidental degeneracy with the quadratic (and
possibly even higher order terms) vanishing identically, we would
have to develop an expansion such as

\begin{eqnarray}
{}^{(2)} g_{ab}(x,n) &=& 
{}^{(2)} g_{ab}(x,0) + 
n \left.{\partial\left[{}^{(2)} g_{ab}(x,n)\right]
\over \partial n}\right|_{n=0} 
\nonumber\\
&& \qquad 
+ {n^{2N}\over{(2N)!}} \;
\left.{\partial^{2N}\left[{}^{(2)} g_{ab}(x,n)\right]
\over (\partial n)^{2N}}\right|_{n=0} +
O[n^{2N+1}].
\end{eqnarray}

\noindent
Applied to the metric determinant this implies an expansion such as

\begin{eqnarray}
\label{E-k-n}
\sqrt{{}^{(2)} g(x,n)} &=& 
\sqrt{{}^{(2)} g(x,0)} 
\left( 
1 + {n^{2N} \; k_N(x) \over {(2N)!}}  + O[n^{2N+1}] 
\right).
\end{eqnarray}

\noindent
Where $k_N(x)$ denotes the first non-zero sub-dominant term 
in the above expansion, and we know by explicit construction that
\begin{eqnarray}
k_N(x) 
&=&
+{1\over2}
\tr \left(\left.
{\partial^{2N}\left[{}^{(2)} g_{ab}(x,n)\right] \over (\partial n)^{2N}}
\right|_{n=0}\right)
\\
&=& 
-\tr \left(\left.  
\frac{\partial^{2N-1} K_{ab}(x,n)}{(\partial n)^{2N-1}}
\right|_{n=0}\right) 
\\
&=& 
-\left(\left. 
\frac{\partial^{2N-1} K(x,n)}{(\partial n)^{2N-1}}
\right|_{n=0}\right),
\end{eqnarray}
since the trace is taken with ${}^{(2)} g^{ab}(x,0) $ and this
commutes with the normal derivative.  We know that the first non-zero
subdominant term in the expansion (\ref{E-k-n}) must be of even
order in $n$ ({\em i.e.} $n^{2N}$), and cannot correspond to an
odd power of $n$, since otherwise the throat would be a point of
inflection of the area, not a minimum of the area. Furthermore,
since $k_N(x)$ is by definition non-zero the flare-out condition
must be phrased as the constraint $k_N(x) > 0$, with this now being
a strict inequality. More formally, this leads to the definition
below.

\noindent
{\bf Definition: $N$-fold degenerate flare-out condition:}\\ {\em
A two surface satisfies the ``$N$-fold degenerate flare-out''
condition at a point $x$ if and only if it is extremal, $\tr(K)=0$,
if in addition the first $2N-2$ normal derivatives of the trace of
the extrinsic curvature vanish at $x$, and if finally at the point
$x$ one has}

\begin{equation}
{\partial^{2N-1}\tr(K)\over (\partial n)^{2N-1} } < 0,
\end{equation}

\noindent
{\em where the inequality is strict. (In the previous notation this
is equivalent to the statement that $k_N(x) > 0$.) }

Physically, at an $N$-fold degenerate point, the wormhole
throat is seen to be extremal up to order $2N-1$ with respect to
normal derivatives of the metric, i.e., the flare-out property is
delayed spatially with respect to throats in which the flare-out
occurs at second order in $n$. The way we have set things up,  the
$1$-fold degenerate flare-out condition is completely equivalent
to the strong flare-out condition.

If we now consider the extrinsic curvature directly we see, by
differentiating (\ref{E-k-n}), first that

\begin{eqnarray}
K(x,n) &=&  - {n^{2N-1} \; k_N(x) \over(2N-1)!}  + O[n^{2N}],
\end{eqnarray}

\noindent
and secondly that

\begin{eqnarray}
{\partial K(x,n) \over\partial n} &=& 
- {n^{2N-2} \; k_N(x) \over(2N-2)!} + O[n^{2N-1}].
\end{eqnarray}

\noindent 
{From} the dominant $n\to 0$ behaviour we see that if (at some
point $x$) $2N$ happens to equal $2$, then the flare-out condition
implies that ${\partial K(x,n) /\partial n}$ must be negative {\em
at and near the throat}. This can also be deduced directly from
the equivalent strong flare-out condition: if  ${\partial K(x,n)
/\partial n}$ is negative and non-zero at the throat, then it must
remain negative in some region surrounding the throat.  On the
other hand, if $2N$ is greater than $2$ the flare-out condition
only tells us that ${\partial K(x,n) /\partial n}$ must be negative
{\em in some region surrounding the the throat}, and does not
necessarily imply that it is negative at the throat itself. (It
could merely be zero at the throat.)

Thus for degenerate throats, the flare-out conditions should
be rephrased in terms of the first non-zero normal derivative beyond
the linear term. Analogous issues arise even for Morris--Thorne
wormholes~\cite[page 405, equation (56)]{Morris-Thorne}, see also
the discussion presented in \cite[pages  104--105, 109]{Visser}.
Even if the throat is non-degenerate ($1$-fold degenerate) there
are technical advantages to phrasing the flare-out conditions this
way:  It allows us to put constraints on the extrinsic curvature
near but not on the throat.

\noindent
{\bf Technical point II: Hyperspatial tubes}\\
A second class of wormholes requiring even more technical
fiddles arises when there is a central section which is completely
uniform and independent of $n$. [So that $K_{ab}=0$ over the whole
throat for some finite range $n\in(-n_0,+n_0)$.] This central
section might be called a ``hyperspatial tube''. The flare-out
condition should then be rephrased as stating that whenever extrinsic
curvature first deviates from zero [at some point $(x,\pm n_0)$]
one must formulate constraints such as

\begin{equation}
\left.{\partial\tr(K) \over \partial n}\right|_{\pm n_0^\pm} \leq 0. 
\end{equation}

In this case $\tr(K)$ is by definition not an analytic function of
$n$ at $n_0$, so the flare-out constraints have to be interpreted
in terms of one-sided derivatives in the region outside the
hyperspatial tube.  [That is, we are concerned with the possibility
that $\sqrt{g(x,n)}$ could be constant for $n< n_0$ but behave as
$(n-n_0)^{2N}$ for $n>n_0$. In this case derivatives, at $n=n_0$,
do not exist beyond order $2N$.]

\section{Geometry of a generic static throat}

Using Gaussian normal coordinates in the region surrounding the throat

\begin{equation}
{}^{(3)}R_{abcd} = {}^{(2)}R_{abcd} - (K_{ac} K_{bd} - K_{ad} K_{bc} ).
\end{equation}

\noindent 
See \cite[page 514, equation (21.75)]{MTW}.
Because two dimensions is special this reduces to:

\begin{equation}
{}^{(3)}R_{abcd} = 
{{}^{(2)}R\over 2} \; (g_{ac} g_{bd} - g_{ad} g_{bc} ) 
                    - (K_{ac} K_{bd} - K_{ad} K_{bc} ).
\end{equation}

\noindent
Of course we still have the standard dimension-independent results
that:

\begin{equation}
{}^{(3)}R_{nabc} =  - (K_{ab:c}  - K_{ac:b}  ).
\end{equation}

\begin{equation}
{}^{(3)}R_{nanb} =  {\partial K_{ab} \over \partial n}  + (K^2)_{ab}.
\end{equation}

\noindent
See \cite[page 514, equation (21.76)]{MTW} and \cite[page
516 equation (21.82)]{MTW}. Here the index $n$ refers to the spatial
direction normal to the two-dimensional throat.

Thus far, these results hold both on the throat and in the region
surrounding the throat: these results hold as long as the Gaussian
normal coordinate system does not break down. (Such breakdown being
driven by the fact that the normal geodesics typically intersect after
a certain distance.) In the interests of notational tractability we
now particularize attention to the throat itself, but shall
subsequently indicate that certain of our results can be extended off
the throat itself into the entire region over which the Gaussian normal
coordinate system holds sway.

Taking suitable contractions, 
{\em and using the extremality condition $\tr(K)=0$},

\begin{eqnarray}
{}^{(3)}R_{ab} &=& 
{{}^{(2)}R\over2} \; g_{ab} 
+  {\partial K_{ab} \over \partial n}  
+ 2 (K^2)_{ab}.
\\
{}^{(3)}R_{na} &=&   -K_{ab}{}^{:b}.
\\
{}^{(3)}R_{nn} &=&  
\tr\left({\partial K \over \partial n}\right)  + \tr(K^2)
\\
&=& {\partial\tr(K)\over\partial n} - \tr(K^2).
\end{eqnarray}

\noindent
So that

\begin{eqnarray}
{}^{(3)}R &=&  
{}^{(2)}R 
+  2 \tr\left({\partial K \over \partial n}\right)  
+ 3\tr(K^2)
\\
&=&  
{}^{(2)}R +  2 {\partial \tr(K) \over \partial n}  - \tr(K^2).
\end{eqnarray}

To effect these contractions, we make use of the decomposition of
the three-space metric in terms of the throat two-metric and the
set of two vectors ${e^{i}_a}$ tangent to the throat and the
three-vector $n^i$ normal to the 2-surface

\begin{equation}
{}^{(2+1)}g^{ij} = e^{i}_a e^{j}_b g^{ab} + n^i n^j.
\end{equation}

For the three-space Einstein tensor ({\em cf.} \cite[page 552]{MTW})
we see

\begin{eqnarray}
{}^{(3)}G_{ab} &=&
{\partial K_{ab} \over \partial n}  + 2 (K^2)_{ab} - 
g_{ab} {\partial \tr(K) \over \partial n} +
{1\over2} g_{ab} \; \tr(K^2).
\\
{}^{(3)}G_{na} &=&   
-K_{ab}{}^{:b}.
\\
{}^{(3)} G_{nn} &=& 
- {1\over2} {}^{(2)} R  - {1\over2} \tr(K^2).
\end{eqnarray}

\noindent
{\em Aside:}
Note in particular that by the flare-out condition ${}^{(3)}R_{nn}
\leq 0$. This implies that the three-space Ricci tensor ${}^{(3)}R_{ij}$
has at least one negative semi-definite eigenvalue everywhere on
the throat. If we adopt the strong flare-out condition then the
three-space Ricci tensor has at least one negative definite eigenvalue
somewhere on the throat.

This decomposition now allows us to write down the various components
of the space-time Einstein tensor. For example

\begin{eqnarray}
{}^{(3+1)}G_{ab} &=& 
 -  \phi_{|ab} - \phi_{|a} \; \phi_{|b}
+ g_{ab} g^{kl} \left[ \phi_{|kl} +  \phi_{|k} \phi_{|l} \right]
\nonumber\\
&&
+{\partial K_{ab} \over \partial n}  + 2 (K^2)_{ab} - 
g_{ab} {\partial \tr(K) \over \partial n} +
{1\over2} g_{ab} \; \tr(K^2)
\nonumber\\
&=&  
8\pi G \; T_{ab}.
\end{eqnarray}

\noindent
But by the definition of the extrinsic curvature, and using the 
Gauss--Weingarten equations,

\begin{eqnarray}
\phi_{|ab} &=& \phi_{:ab} + K_{ab} \; \phi_{|n}.
\\
\phi_{|na} &=&  K_{a}{}^{b} \; \phi_{:b}.
\end{eqnarray}

\noindent
[See, for example, equations (21.57) and (21.63) of~\cite{MTW}.] Thus 

\begin{equation}
g^{kl} \phi_{|kl} = 
g^{ab} \phi_{:ab} +  (g^{ab}K_{ab}) \; \phi_{|n} 
+ \phi_{|nn}.
\end{equation}

\noindent
But remember that $\tr(K)=0$ at the throat, so

\begin{equation}
g^{kl} \phi_{|kl} = 
g^{ab} \phi_{:ab} + \phi_{|nn}.
\end{equation}

This finally allows us to write

\begin{eqnarray}
{}^{(3+1)}G_{ab} 
&=& 
 -  \phi_{:ab} - \phi_{:a} \; \phi_{:b} - K_{ab} \; \phi_{|n}
\nonumber\\
&&
+ g_{ab} \left[ 
 g^{cd} ( \phi_{:cd} +\phi_{:c} \phi_{:d} )
+ \phi_{|nn} + \phi_{|n} \phi_{|n} 
\right]
\nonumber\\
&&
+ {\partial K_{ab} \over \partial n}  + 2 (K^2)_{ab} - 
g_{ab} {\partial \tr(K) \over \partial n} +
{1\over2} g_{ab} \; \tr(K^2)
\nonumber\\
&=&  
8\pi G \; T_{ab}.
\\
{}^{(3+1)}G_{na} 
&=&   
-  K_a{}^b \phi_{:b} - \phi_{|n} \; \phi_{:a}
-  K_{ab}{}^{:b}  
\nonumber\\
&=& 
8\pi G \; T_{na}.
\\
{}^{(3+1)} G_{nn} &=&  g^{cd} 
\left[ \phi_{:cd} + \phi_{:c} \phi_{:d}  \right]
- {1\over2} {}^{(2)} R  - {1\over2} \tr(K^2) 
\nonumber\\
&=& 
-8\pi G \; \tau.
\\
{}^{(3+1)}G_{\hat t a} &=&   0.
\\
{}^{(3+1)}G_{\hat t n} &=&   0.
\\
{}^{(3+1)}G_{\hat t \hat t} &=&  
{{}^{(2)}R\over2} +  
{\partial \tr(K) \over \partial n}  - 
{1\over2} \tr(K^2)
\nonumber\\
&=&
+8\pi G \; \rho.
\end{eqnarray}

\noindent
Here $\tau$ denotes the {\em tension} perpendicular to the wormhole
throat, it is the natural generalization of the quantity considered
by Morris and Thorne, while $\rho$ is simply the energy density at
the wormhole throat.

\section{Constraints on the stress-energy tensor}

We can now derive several constraints on the stress-energy:

---First---

\begin{equation}
\tau = {1\over16\pi G} 
\left[
{}^{(2)} R  +  \tr(K^2)  -2 g^{cd} (\phi_{:cd} + \phi_{:c} \phi_{:d} )
\right].
\end{equation}

\noindent 
(Unfortunately the signs as given are correct. Otherwise we would have
a lovely lower bound on $\tau$. We will need to be a little tricky
when dealing with the $\phi$ terms.)  The above is the generalization
of the Morris--Thorne result that

\begin{equation}
\tau= {1\over8\pi G r_0^2}
\end{equation}

\noindent
at the throat of the special class of model wormholes they
considered. (With MTW conventions ${}^{(2)} R = 2/r_0^2$ for a
two-sphere.) If you now integrate over the surface of the wormhole

\begin{equation}
\int \sqrt{{}^{(2)}g} \; \tau \; d^2x = {1\over16\pi G} 
\left[
4\pi\chi +  \int \sqrt{{}^{(2)}g} 
\left\{\tr(K^2) -2 g^{cd} \phi_{:c} \phi_{:d}\right\} \;d^2 x 
\right].
\end{equation}

\noindent
Here $\chi$ is the Euler characteristic of the throat, while the
$g^{cd} \phi_{:cd}$ term vanishes by partial integration, since the throat
is a manifold without boundary.

---Second---

\begin{equation}
\rho = {1\over16\pi G} 
\left[  
{}^{(2)}R +  2 {\partial \tr(K) \over \partial n} - \tr(K^2)
\right].
\end{equation}

\noindent
The second term is negative semi-definite by the flare-out condition,
while the third term is manifestly negative semi-definite. Thus

\begin{equation}
\rho \leq {1\over16\pi G} \; {}^{(2)}R.
\end{equation}

\noindent
This is the generalization of the Morris--Thorne result that 

\begin{equation}
\rho= {b'(r_0)\over8\pi G r_0^2} \leq  {1\over8\pi G r_0^2}
\end{equation}
 
\noindent
at the throat of the special class of model wormholes they considered.
(See~\cite[page 107]{Visser}.)

Note in particular that if the wormhole throat does not have the
topology of a sphere or torus then there {\em must} be places on the
throat such that ${}^{(2)}R < 0$ and thus such that $\rho < 0$.  Thus
wormhole throats of high genus will always have regions that violate
the weak and dominant energy conditions. (The simple flare-out
condition is sufficient for this result. For a general discussion of
the energy conditions see~\cite{Visser} or~\cite{Hawking-Ellis}.)

If the wormhole throat has the topology of a torus then it will
generically violate the weak and dominant energy conditions; only
for the very special case ${}^{(2)}R = 0$, $K_{ab}=0$, $\partial
\tr(K) /\partial n =0$ will it possibly satisfy (but still be on the
verge of violating) the weak and dominant energy conditions. This is
a particular example of a degenerate throat in the sense discussed
previously.

Wormhole throats with the topology of a sphere will, provided they
are convex, at least have positive energy density, but we shall
soon see that other energy conditions are typically violated.

If we now integrate over the surface of the wormhole

\begin{equation}
\int \sqrt{{}^{(2)}g} \; \rho \; d^2x = {1\over16\pi G} 
\left[
4 \pi\chi +  
\int \sqrt{{}^{(2)}g} 
\left\{2 {\partial \tr(K) \over \partial n} - \tr(K^2)\right\} \; d^2 x 
\right].
\end{equation}

\noindent
So for a throat with the topology of a torus ($\chi = 0$) 
the simple flare-out
condition yields

\begin{equation}
\int \sqrt{{}^{(2)}g} \; \rho \; d^2x \leq 0,
\end{equation}

\noindent
while the strong or weak flare-out conditions yield

\begin{equation}
\int \sqrt{{}^{(2)}g} \; \rho \; d^2x < 0,
\end{equation}

\noindent
guaranteeing violation of the weak and dominant energy conditions.
For a throat with higher genus topology $(\chi = 2 - 2g)$ 
the simple flare-out condition
is sufficient to yield

\begin{equation}
\int \sqrt{{}^{(2)}g} \; \rho \; d^2x \leq {\chi\over 4 G} < 0 . 
\end{equation}

---Third---

\begin{equation}
\rho-\tau = {1\over16\pi G} 
\left[   
+2 {\partial \tr(K) \over \partial n}
-2 \tr(K^2) 
+2 g^{cd}(\phi_{:cd}  + \phi_{:c} \phi_{:d} )
\right].
\end{equation}

\noindent
Note that the two-curvature ${}^{(2)}R$ has conveniently dropped out
of this equation.  As given, this result is valid only on the
throat itself, but we shall soon see that a generalization can
be constructed that will also hold in the region surrounding the
throat. The first term is negative semi-definite by the simple
flare-out condition (at the very worst when integrated over the throat
it is negative by the weak flare-out condition). The second term is
negative semi-definite by inspection. The third term integrates to
zero though it may have either sign locally on the throat. The fourth
term is unfortunately positive semi-definite on the throat which
prevents us from deriving a truly general energy condition violation
theorem without additional information.

Now because the throat is by definition a compact two surface, we know
that $\phi(x^a)$ must have a maximum somewhere on the throat. At the
global maximum (or even at any local maximum) we have $\phi_{:a}=0$
and $g^{ab} \phi_{:ab}\leq 0$, so at the maxima of $\phi$ one has

\begin{equation}
\rho-\tau \leq 0.
\end{equation}

Generically, the inequality will be strict, and generically there
will be points on the throat at which the null energy condition is
violated.

Integrating over the throat we have

\begin{eqnarray}
&&\int \sqrt{{}^{(2)}g} \; [\rho-\tau] \; d^2 x 
\nonumber\\
&& \qquad
= {1\over16\pi G} 
\int \sqrt{{}^{(2)}g} \left[   
+2 {\partial \tr(K) \over \partial n}
-2 \tr(K^2) 
+2 g^{cd}( \phi_{:c} \phi_{:d} )
\right] d^2x.
\end{eqnarray}

Because of the last term we must be satisfied with the result

\begin{equation}
\int \sqrt{{}^{(2)}g} \; [\rho-\tau] \; d^2 x \leq 
\int \sqrt{{}^{(2)}g} \left[   
2 g^{cd}( \phi_{:c} \phi_{:d} )
\right] d^2x.
\end{equation}

---Fourth---

We can rewrite the difference $\rho-\tau$ as

\begin{equation}
\rho-\tau = {1\over16\pi G} 
\left[   
+2 {\partial \tr(K) \over \partial n}
-2 \tr(K^2) 
+2 \exp(-\phi) \;\; {}^{(2)}\Delta \exp(+\phi)
\right].
\end{equation}

\noindent
So if we multiply by $\exp(+\phi)$ before integrating, the
two-dimensional Laplacian  ${}^{(2)}\Delta$
vanishes by partial integration and we have

\begin{eqnarray}
\label{E-t-anec}
&&\int \sqrt{{}^{(2)}g} \; \exp(+\phi) \; [\rho-\tau] \; d^2 x 
\nonumber\\
&& \qquad 
= {1\over8\pi G} 
\int \sqrt{{}^{(2)}g} \; \exp(+\phi) \; 
\left[   
+{\partial \tr(K) \over \partial n}
-\tr(K^2) 
\right] d^2x.
\end{eqnarray}

Thus the strong flare-out condition (or less restrictively, the weak
$e^\phi$--weighted flare-out condition) implies the violation of this
``transverse averaged null energy condition'' (the NEC averaged over 
the throat)

\begin{equation}
\int \sqrt{{}^{(2)}g} \; \exp(+\phi) \; [\rho-\tau] \; d^2 x < 0.
\end{equation}

---Fifth---

We can define an average transverse pressure on the throat by

\begin{eqnarray}
\bar p &\equiv& {1\over16\pi G} \; \; g^{ab} \;\; {}^{(3+1)}G_{ab}
\\
&=&
{1\over16\pi G}  \left[
g^{cd} ( \phi_{:cd} +\phi_{:c} \phi_{:d} )
+ 2 \phi_{|nn} + 2 \phi_{|n} \phi_{|n} 
- {\partial \tr(K) \over \partial n} + \tr(K^2)
\right].
\nonumber\\
&&
\end{eqnarray}

\noindent
The last term is manifestly positive semi-definite, the penultimate
term is positive semi-definite by the flare-out condition. The
first and third terms are of indefinite sign while the second and
fourth are also positive semi-definite. Integrating over the surface
of the throat

\begin{eqnarray}
\int \sqrt{{}^{(2)}g} \; \bar p  \; d^2x &\geq&
{1\over8\pi G}
\int \sqrt{{}^{(2)}g}  \; \phi_{|nn} \; d^2 x.
\end{eqnarray}

A slightly different constraint, also derivable from the above, is

\begin{eqnarray}
\int \sqrt{{}^{(2)}g} \; e^\phi \; \bar p  \; d^2x &\geq&
{1\over8\pi G}
\int \sqrt{{}^{(2)}g}  \; (e^\phi)_{|nn} \; d^2 x.
\end{eqnarray}

These inequalities relate transverse pressures to normal derivatives of
the gravitational potential. In particular, if the throat lies at a
minimum of the gravitational red-shift the second normal derivative
will be positive, so the transverse pressure (averaged over the
wormhole throat) must be positive.

---Sixth---

Now look at the quantities $\rho-\tau+2\bar p$ and $\rho-\tau-2\bar
p$. We have

\begin{eqnarray}
\rho-\tau+2\bar p
&=& 
{1\over4\pi G} \left\{
g^{cd} ( \phi_{:cd} +\phi_{:c} \phi_{:d} )
+  \phi_{|nn} +  \phi_{|n} \phi_{|n} 
\right\}
\\
&=& 
{1\over4\pi G} \left\{
g^{ij} ( \phi_{|ij} +\phi_{|i} \phi_{|j} )
\right\}.
\end{eqnarray}

This serves as a nice consistency check. The combination of
stress-energy components appearing above is equal to $\rho + g^{ij}
T_{ij}$ and is exactly that relevant to the strong energy
condition. See equations
(\ref{E-static-SEC-b})---(\ref{E-static-SEC-e}).  See also equations
(\ref{E-static-stress-energy-b})---(\ref{E-static-stress-energy-e}).
Multiplying by $e^\phi$ and integrating

\begin{eqnarray}
\int \sqrt{{}^{(2)}g} \; e^\phi \; \left[\rho-\tau+2\bar p\right]  \; d^2x =
{1\over4\pi G} \int \sqrt{{}^{(2)}g}  \; (e^\phi)_{|nn} \; d^2 x.
\end{eqnarray}

This relates this transverse integrated version of the strong energy
condition to the normal derivatives of the gravitational potential.

On the other hand

\begin{equation}
\rho-\tau-2\bar p
=
{1\over4\pi G} \left\{
- \phi_{|nn} -  \phi_{|n} \phi_{|n} 
+ {\partial\tr(K)\over\partial n} - \tr(K^2)
\right\}.
\end{equation}

The second and fourth terms are negative semi-definite, while the
third term is negative semi-definite by the flare-out condition.

---Summary---

There are a number of powerful constraints that can be placed on the
stress-energy tensor at the wormhole throat simply by invoking the
minimality properties of the wormhole throat. Depending on the precise
form of the assumed flare-out condition, these constraints give the
various energy condition violation theorems we are seeking. Even under
the weakest assumptions (appropriate to a degenerate throat) they
constrain the stress-energy to at best be on the verge of violating
the various energy conditions.

\section{Special case: The isopotential throat}

Suppose we take $\phi_{:a}=0$. This additional constraint corresponds
to asserting that the throat is an {\em isopotential} of the
gravitational red-shift. In other words, $\phi(n,x^a)$ is simply
a constant on the throat.  For instance, all the Morris--Thorne
model wormholes~\cite{Morris-Thorne} possess this symmetry. Under
this assumption there are numerous simplifications.

We will not present anew all the results for the Riemann curvature
tensor but instead content ourselves with the Einstein tensor

\begin{eqnarray}
{}^{(3+1)}G_{ab} &=& 
+ g_{ab} \left( 
\phi_{|nn}  + \phi_{|n} \phi_{|n} 
\right) - K_{ab} \; \phi_{|n}
\nonumber\\
&&  
+ {\partial K_{ab} \over \partial n}  + 2 (K^2)_{ab} - 
g_{ab} {\partial \tr(K) \over \partial n} +
{1\over2} g_{ab} \; \tr(K^2)
\nonumber\\
&=&  
8\pi G \; T_{ab}.
\\
{}^{(3+1)}G_{na} &=&   
-K_{ab}{}^{:b}  = 
8\pi G \; T_{na}.
\\
{}^{(3+1)} G_{nn} &=&  
- {1\over2} {}^{(2)} R  - {1\over2} \tr(K^2) = 
-8\pi G \; \tau.
\\
{}^{(3+1)}G_{\hat t a} &=&   0.
\\
{}^{(3+1)}G_{\hat t n} &=&   0.
\\
{}^{(3+1)}G_{\hat t \hat t} &=&  
{{}^{(2)}R\over2} +  
{\partial \tr(K) \over \partial n}- {1\over2} \tr(K^2)
= +8\pi G \; \rho.
\end{eqnarray}

\noindent
Thus for an isopotential throat

\begin{eqnarray}
\tau &=& {1\over16\pi G} 
\left[
{}^{(2)} R  +  \tr(K^2)  
\right] 
\geq {1\over16\pi G} {}^{(2)} R.
\\
\rho &=& {1\over16\pi G} 
\left[  
{}^{(2)}R +  2 {\partial \tr(K) \over \partial n} - \tr(K^2)
\right]  \leq {1\over16\pi G} {}^{(2)} R.
\\
\rho-\tau &=& {1\over16\pi G} 
\left[   
+2 {\partial \tr(K) \over \partial n}
-2 \tr(K^2) 
\right] \leq 0.
\end{eqnarray}

\noindent
This  gives us a very powerful result: using only the simple
flare-out condition, the NEC is on the verge of being violated
everywhere on an isopotential throat.

By invoking the strong flare-out condition the  NEC is definitely
violated somewhere on an isopotential throat.

Invoking the weak flare-out condition we can still say that the
surface integrated NEC is definitely  violated on an isopotential
throat.

\section{Special case: The extrinsically flat throat}

Suppose now that we take $K_{ab}=0$. This is a much stronger
constraint than simple minimality of the area of the wormhole throat
and corresponds to asserting that the three-geometry of the throat
is (at least locally) symmetric under interchange of the two regions
it connects.  For instance, all the Morris--Thorne model
wormholes~\cite{Morris-Thorne} possess this symmetry and have
throats that are extrinsically flat.  Under this assumption there
are also massive simplifications. (Note that we are not making the
isopotential assumption at this stage.)

Again, we will not present all the results but content ourselves
with the Einstein tensor

\begin{eqnarray}
{}^{(3+1)}G_{ab} &=& 
 -  \phi_{:ab} - \phi_{:a} \; \phi_{:b}
+ g_{ab} \left[ 
 g^{cd} ( \phi_{:cd} +\phi_{:c} \phi_{:d} )
+ \phi_{|nn}  + \phi_{|n} \phi_{|n} 
\right]
\nonumber\\
&&+ {\partial K_{ab} \over \partial n}  - 
g_{ab} {\partial \tr(K) \over \partial n} 
\nonumber\\
&=&  
8\pi G \; T_{ab}.
\\
{}^{(3+1)}G_{na} &=&    
- \phi_{|n} \; \phi_{|a}  = 
8\pi G \; T_{na}.
\\
{}^{(3+1)} G_{nn} &=&  g^{cd} 
\left[ \phi_{:cd} + \phi_{:c} \phi_{:d}  \right]
- {1\over2} {}^{(2)} R  = 
-8\pi G \; \tau.
\\
{}^{(3+1)}G_{\hat t a} &=&   0.
\\
{}^{(3+1)}G_{\hat t n} &=&   0.
\\
{}^{(3+1)}G_{\hat t \hat t} &=&  
{{}^{(2)}R\over2} +  {\partial \tr(K) \over \partial n}
= +8\pi G \; \rho.
\end{eqnarray}

Though the stress-energy tensor is now somewhat simpler than the
general case, the presence of the $\phi_{:a}$ terms precludes the
derivation of any truly new general theorems.

\section{Special case: The extrinsically flat isopotential throat}

Finally, suppose we take both $K_{ab}=0$ and $\phi_{:a}=0$.  A
wormhole throat that is both extrinsically flat and isopotential is
particularly simple to deal with, even though it is still much more
general than the Morris--Thorne wormhole.  Once again, we will not
present all the results but content ourselves with the Einstein tensor

\begin{eqnarray}
{}^{(3+1)}G_{ab} &=& 
+ g_{ab} \left( 
\phi_{|nn} + \phi_{|n} \phi_{|n} 
\right)
+ {\partial K_{ab} \over \partial n}  - 
g_{ab} {\partial \tr(K) \over \partial n}
\nonumber\\
&=&  
8\pi G \; T_{ab}.
\\
{}^{(3+1)}G_{na} &=&  0. 
\\
{}^{(3+1)} G_{nn} &=&  
- {1\over2} {}^{(2)} R  = 
-8\pi G \; \tau.
\\
{}^{(3+1)}G_{\hat t a} &=&   0.
\\
{}^{(3+1)}G_{\hat t n} &=&   0.
\\
{}^{(3+1)}G_{\hat t \hat t} &=&  
{{}^{(2)}R\over2} +  
{\partial \tr(K) \over \partial n}
= +8\pi G \; \rho.
\end{eqnarray}

\noindent
In this case $\rho-\tau$ is particularly simple:

\begin{equation}
\rho - \tau = {1\over8\pi G} {\partial \tr(K) \over \partial n}.
\end{equation}

\noindent
This quantity  is manifestly negative semi-definite by the simple
flare-out condition.

For the strong flare-out condition we deduce that the NEC must be
violated somewhere on the wormhole throat.

Even for the weak flare-out condition we have

\begin{equation}
\int \sqrt{{}^{(2)}g} \; \left[ \rho - \tau \right] \; d^2 x < 0.
\end{equation}

We again see that generic violations of the null energy condition
are the rule.

\section{The region surrounding the throat}

Because the spacetime is static, one can unambiguously define the
energy density everywhere in the spacetime by setting

\begin{equation}
\rho = { {}^{(3+1)} G_{\hat t\hat t}\over8\pi G}.
\end{equation}

The normal tension, which we have so far defined only on the wormhole
throat itself, can meaningfully be extended to the entire region where
the Gaussian normal coordinate system is well defined by setting

\begin{equation}
\tau = - { {}^{(3+1)} G_{nn}\over8\pi G}.
\end{equation}

Thus in particular

\begin{equation}
\rho-\tau 
= { {}^{(3+1)} G_{\hat t\hat t} + {}^{(3+1)} G_{nn}\over8\pi G} 
= { {}^{(3+1)} R_{\hat t\hat t} + {}^{(3+1)} R_{nn}\over8\pi G},
\end{equation}

\noindent
with this quantity being well defined throughout the Gaussian normal
coordinate patch. (The last equality uses the fact that $g_{\hat
t\hat t}=-1$ while $g_{nn}=+1$.)  But we have already seen how to
evaluate these components of the Ricci tensor. Indeed

\begin{eqnarray}
{}^{(3+1)}R_{\hat t\hat t} &=&  g^{ij}
\left[  
\phi_{|ij} + \phi_{|i} \phi_{|j}
\right].
\\
{}^{(3+1)}R_{nn} &=&  {}^{(3)}R_{nn} -
\left[  
\phi_{|nn} + \phi_{|n} \phi_{|n}
\right]
\\
&=& {\partial\tr(K)\over\partial n} - \tr(K^2) -
\left[  
\phi_{|nn} + \phi_{|n} \phi_{|n}
\right],
\end{eqnarray}

\noindent
where we have been careful to {\em not} use the extremality condition
$\tr(K)=0$. Therefore

\begin{eqnarray}
\rho-\tau
&=& 
{1\over8\pi G}  
\left[ 
{\partial\tr(K)\over\partial n} - \tr(K^2) + 
g^{ab} 
\left(  
\phi_{|ab} + \phi_{|a} \phi_{|b}
\right)
\right]
\\
&=& 
{1\over8\pi G}  
\left[ 
{\partial\tr(K)\over\partial n} - \tr(K^2) + \tr(K) \phi_{|n} +
g^{ab} 
\left(  
\phi_{:ab} + \phi_{:a} \phi_{:b}
\right)
\right],
\nonumber\\
&&
\end{eqnarray}

\noindent
where in the last line we have used the Gauss--Weingarten equations.

If the throat is {\em isopotential}, where isopotential now means
that near the throat the surfaces of constant gravitational potential
coincide with the surfaces of fixed $n$, this simplifies to:

\begin{eqnarray}
\rho-\tau
&=& 
{1\over8\pi G}  
\left[ 
{\partial\tr(K)\over\partial n} - \tr(K^2) + \tr(K) \phi_{|n}
\right].
\end{eqnarray}

\noindent
If the throat is non-degenerate and satisfies the simple
flare-out condition, then at the throat the first and second terms
are negative semi-definite,  and the third is zero. Then the null
energy condition is either violated or on the verge of being violated
at the throat. 

If the throat is non-degenerate and satisfies the strong
flare-out condition at the point $x$, then  the first term is
negative definite, the second is negative semi-definite, and the
third is zero. Then the null energy condition is violated at the
point $x$ on the throat.  

If the throat satisfies the $N$-fold degenerate flare-out
condition at the point $x$, then by the generalization of the
flare-out conditions applied to degenerate throats the first term
will be $O[n^{2N-2}]$ and negative definite in some region surrounding
the throat. The second term is again negative semi-definite.  The
third term can have either sign but will be $O[n^{2N-1}]$. Thus
there will be some region $n\in(0,n_*)$ in which the first term
dominates.  Therefore the  null energy condition is violated along
the line $\{x\}\times(0,n_*)$. If at every point $x$ on the throat
the $N$-fold degenerate flare-out condition is satisfied for some
{\em finite} $N$, then there will be an open region surrounding
the throat on which the null energy condition is everywhere violated.

This is the closest one can get in generalizing to arbitrary
wormhole shapes the discussion on page 405 [equation (56)] of
Morris--Thorne~\cite{Morris-Thorne}.  Note carefully their use of
the phrase  ``at or near the throat''. In our parlance, they are
considering a spherically symmetric extrinsically flat isopotential
throat that satisfies the $N$-fold degenerate flare-out condition
for some finite but unspecified $N$. See also page 104, equation
(11.12) and page 109, equation (11.54) of~\cite{Visser}, and
contrast this with equation (11.56).

If the throat is not isopotential we multiply by $\exp(\phi)$ and
integrate over surfaces of constant $n$. Then

\begin{eqnarray}
&&\int \sqrt{{}^{(2)} g} \; \exp(\phi) \; [\rho-\tau] \;  d^2x =
\nonumber\\
&&\qquad
{1\over8\pi G}  \int \sqrt{{}^{(2)} g} \; \exp(\phi)
\left[ 
{\partial\tr(K)\over\partial n} - \tr(K^2) + \tr(K) \phi_{|n}
\right]
 d^2x.
\nonumber\\
\end{eqnarray}

This generalizes the previous version (\ref{E-t-anec}) of the
transverse averaged null energy condition to constant $n$ hypersurfaces
near the throat.  For each point $x$ on the throat, assuming the
$N$-fold degenerate flare-out condition,  we can by the previous
argument find a range of values [$n\in(0,n_*(x))$] that will make
the integrand negative. Thus there will be a set of values of $n$
for which the integral is negative.  Again we deduce violations of
the null energy condition.

\section{Discussion}

We have presented a definition of a wormhole throat that is much
more general than that of the Morris--Thorne wormhole~\cite{Morris-Thorne}.
The present definition works well in any static spacetime and nicely
captures the essence of the idea of what we would want to call a
wormhole throat.

We do not need to make any assumptions about the existence of any
asymptotically flat region, nor do we need to assume that the
manifold is topologically non-trivial. It is important to realise
that the essence of the definition lies in the geometrical structure
of the wormhole throat.

Starting from our definition we have used the theory of embedded
hypersurfaces to place restrictions on the Riemann tensor and
stress-energy tensor at the throat of the wormhole. We find, as
expected, that the wormhole throat generically violates the null
energy condition and we have provided several theorems regarding
this matter. These theorems generalise the Morris--Thorne results
on exotic matter~\cite{Morris-Thorne}, and are complementary to
the topological censorship theorem~\cite{Topological-censorship}

Generalization to the time dependent situation is clearly of
interest. Unfortunately we have encountered many subtleties of
definition, notation, and formalism in this endeavour. We defer the
issue of time dependent wormhole throats to a future publication.

\section*{Acknowledgements}

M.V. wishes to gratefully acknowledge the hospitality shown during
his visits to the Laboratory for Space Astrophysics and Fundamental
Physics (LAEFF, Madrid). This work was supported in part by the US
Department of Energy (M.V.) and by the Spanish Ministry of Science
and Education (D.H.).



\end{document}